\documentclass{ws-ijbc}
\usepackage{ws-rotating}     
\usepackage{graphicx}
\usepackage{epstopdf}
\begin{document}

\catchline{}{}{}{}{} 

\markboth{Maistrenko  {\it et al.}}{Cascades of multi-headed chimera states  for coupled phase oscillators}

\title{CASCADES OF MULTI-HEADED CHIMERA STATES FOR COUPLED PHASE OSCILLATORS}

\author{YURI L.~MAISTRENKO$^{\star \dagger}$, ANNA VASYLENKO$^{\dagger \ddagger}$, OLEKSANDR SUDAKOV$^{\dagger \amalg}$,\\ ROMAN LEVCHENKO$^{\dagger}$ and VOLODYMYR L.~MAISTRENKO$^{\dagger}$ }

\address{$^{\star}$Institute of Mathematics, National Academy of Sciences of Ukraine, \\  
Tereshchenkivska st.~3, 01030, Kyiv, Ukraine, y.maistrenko@biomed.kiev.ua \\
$^{\dagger}$National Scientific Centre for Medical and Biotechnical Research,\\ National Academy of Sciences of Ukraine, \\Volodymyrska Str.~54, 01030, Kyiv, Ukraine, maistren@nas.gov.ua \\
$^{\ddagger}$University of Antwerp, Groenenborgerlaan 171, BE-2020, \\
Antwerp, Belgium, anna.vasylenko@uantwerpen.be\\
$^{\amalg}$ Taras Shevchenko National University of Kyiv, Volodymyrska Str.~60,\\ 01030, Kyiv, Ukraine, saa@grid.org.ua
}

\maketitle

\begin{history}
\received{(to be inserted by publisher)}
\end{history}

\begin{abstract}
\textit{Chimera state} is a recently discovered dynamical phenomenon in arrays of nonlocally
coupled oscillators, that displays a self-organized spatial pattern of co-existing coherence and incoherence.
We discuss the appearance of the chimera states in networks of phase oscillators with attractive and with repulsive interactions, 
i.e. when the coupling respectively favors synchronization or works against it. 
By systematically analyzing the dependence of the spatiotemporal dynamics on the level of coupling attractivity/repulsivity and the range of coupling, we uncover that different types of chimera states exist in wide domains of the parameter space as cascades of the states with increasing
number of intervals of irregularity, so-called \textit{chimera's heads}. 
We report three scenarios for the chimera birth:  1) via saddle-node bifurcation on a resonant invariant circle, also known as SNIC or SNIPER, 2) via blue-sky catastrophe, when two periodic orbits, stable and saddle, approach each other creating a saddle-node periodic orbit, and 3) via homoclinic transition with complex multistable dynamics including an ''eight-like'' limit cycle resulting eventually in a chimera state.
\end{abstract}


\begin{twocolumn}
\section{Introduction}
\noindent Chimera state represents a remarkable spatiotemporal pattern where synchronized and phase-locked oscillators coexist with desynchronized and incoherent ones. Dynamically, it represents a sort of spatially extended symmetry breaking which develops in arrays of identical oscillators, moreover without any sign of asymmetry or external excitation.  Certainly, it is a robust manifestation of strongly nonlinear nature of self-organization in homogeneous systems.  

This kind of coherent-incoherent behavior was first observed ten years ago for  nonlocally coupled complex
Ginzburg-Landau equation and for its phase approximation model~\cite{kb2002}
\begin{equation}
\label{model}
\dot{\varphi}_i = \omega +
\frac{K}{N}\sum\limits_{j=1}^N G_{ij}\sin(\varphi_j-\varphi_i-\alpha), \\ ~~~i=1,\ldots,N,
\end{equation}
with coupling function $G$ decaying exponentially with the distance between the oscillators:
\begin{equation}
G_{\exp}(x,k)=\frac{k}{2}\exp(-kx),
\label{modelGexp}
\end{equation}
where $x=\frac{|j-i|}{N}$. The term \textit{chimera state} was introduced two years later in ~\cite{as2004}, where model (\ref{model}) was analyzed with the cosine coupling function
\begin{equation}
G_{\cos}(x,A)=1+A\cos2\pi{x}.
\label{modelGcos}
\end{equation}
Since that, chimera phenomenon has got a lot of attention and has been studied for different types of oscillatory networks, with various types of non-local interaction, and by different methods including both finite- and infinite-dimensional approaches ~\cite{sk2004,as2006,s2006,amsw2008,omt2008,ssa2008,l2009,mls2010,owm2010,bpr2010,woym2011,lroa2011,wo2011,
omhs2011,lrk2012,zlzy2012,
owyms2012,pa2013,hmrhos2012,tns2012,nts2013,lpm2013,mtfh2013,oohs2013}.

In the model (\ref{model}), the oscillators $\varphi_i, i=1,...,N,$ are assumed to be uniformly distributed over a one-dimensional ring (of unit length, for definiteness), so that the index $i$ is periodic mod $N$. $\omega$ denotes the natural frequency of the oscillators that can be set to zero, $K$ is coupling coefficient that can be set to one. An important role for the system dynamics is played by the phase lag parameter $\alpha$ . Indeed, in the most of the papers listed above, chimera state was found for special values of parameter $\alpha$ which are smaller but close to $\pi/2$.  E.g., classical Kuramoto parameter values at which chimera state was first detected are $\alpha=1.457, k=4.0$; Strogatz 2004 values are  $\alpha=\pi/2-0.18, A=0.995$. Note, that in both cases chimera state co-exists  with the fully synchronized state $\varphi_i=...=\varphi_N$ (which is indeed stable for all $-\pi/2<\alpha<\pi/2$). Due to the co-existence,  chimera states can be obtained in numerical simulations only from specially prepared initial conditions ~\cite{as2006}.

Recently, chimera state was reported for Eq.(\ref{model}) with the simplest possible form
of nonlocal coupling given by a step function ~\cite{owm2010,woym2011,wo2011,omhs2011,orhms2012,owyms2012,nts2013,oohs2013}:
\begin{equation}
G_{step}(x,r)=\left\{
\begin{array}{lcl}
1/2r,&\mbox{ if }& |x|\le r,\\[2mm]
0,&\mbox{ if }& |x|>r.
\end{array}
\right.
\label{Coupling:PW}
\end{equation}
In this case, each oscillator $\varphi_i$ is coupled with equal strength to its $P$ nearest neighbors on either side, and the model (\ref{model}) obeys the form:
\begin{equation}
\label{model1}
\dot{\varphi_i} = \omega+\frac{K}{2P}\sum\limits_{j=-P}^{P}\sin(\varphi_{i+j} - \varphi_i- \alpha).
\end{equation}
An essential parameter in Eq.(\ref{model1}) is \textit{radius of coupling} $r=P/N$.  At $r=1/N$ coupling in the model is local (nearest-neighbor), and for $r=0.5$ it is global (mean-field). Our target is an intermediate case $1/N<r<0.5$  when chimeras can indeed exist. 


\begin{figure*}[ht!]
  \center{\includegraphics[width=1\linewidth]{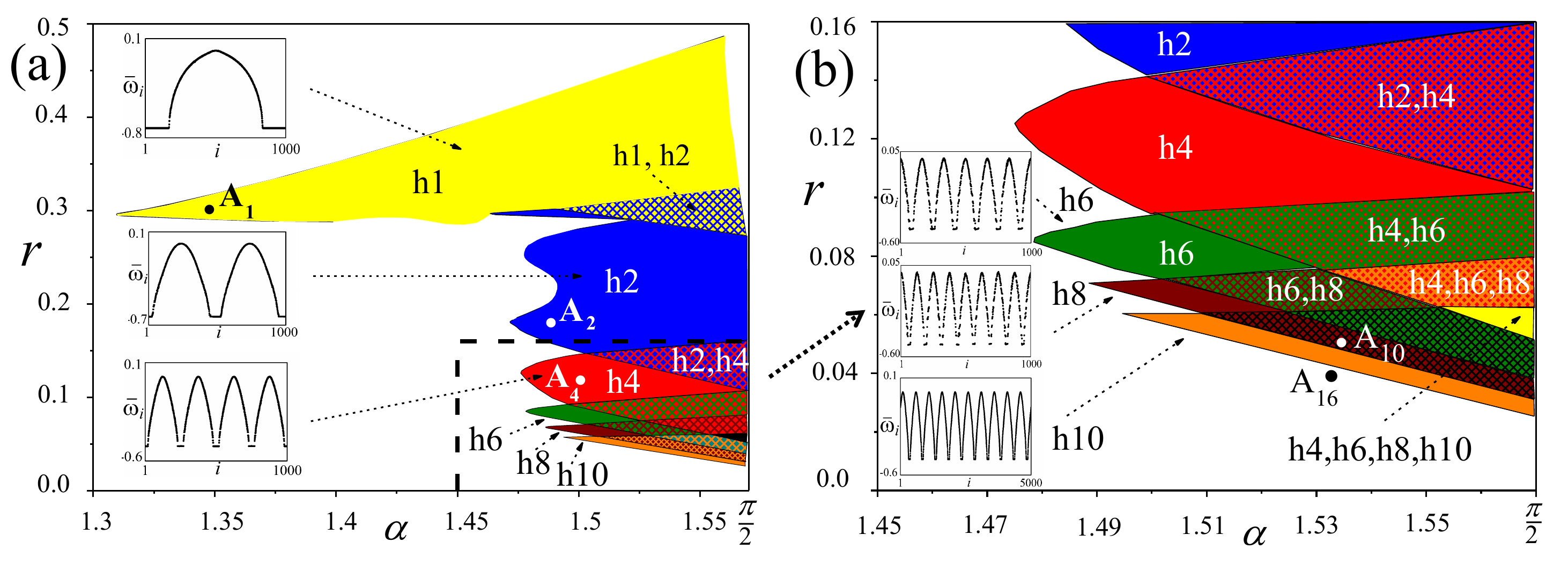}}
 \caption{(a) Parameter regions of multi-headed chimera states in Kuramoto model (\ref{model1}) with $N=1000$ oscillators,. (b) Enlargement of the rectangle from (a).  Insets show typical profiles
of average frequencies of individual oscillators in parameter points indicated by arrows.  Hatched
regions denote multistable regimes where two or more chimera states coexist.  Calculations where performed up to $T=10000$ time units, average frequencies were calculated for $\Delta$$T=100$,  grid $h\alpha=hr=0.001$.}
  \label{f1}
\end{figure*}


As it was found in ~\cite{owm2010}, chimera states are not fixed objects in the space but they wander along the ring performing some kind of random walk, and with essentially hyperchaotic dynamics at finite time scales ~\cite{woym2011}. Amplitude of the chimera's wandering grows with decrease of the system size $N$ so that chimera collapses falling on the coherent state, which can be easily observed at N  between 40 and 50. On the other hand, as $N$ increases, chimera's life time grows exponentially ~\cite{wo2011} so that for $N$ around 100 or more only very long calculations of Eq.(\ref{model1}) are expected to confirm the transient nature of the chimera state. Therefore, for $N$ large enough, chimeras can be considered as \textit{effective attractors} of nonlocally coupled phase oscillators.

In the Paper, we study appearance of chimera states for coupled phase oscillators given by Eq.(\ref{model1}) with both \textit{attractive} (Ch.2) and  \textit{repulsive} (Ch.3) coupling.
Our numerics are based on the Runge-Kutta solver DOPRI5
that has been integrated by software for large nonlinear
dynamical networks ~\cite{sls2011}, allowing for parallelized simulations
with different sets of parameters and initial conditions.
  With use of massive numerical simulations we find that chimeras exist in wide domains of the parameter
space as cascades of the states with increasing number of regions of irregularity, so-
called \textit{chimera's heads}.  The measure of coupling  attractivity/repulsivity in Eq.(\ref{model1}) is controlled by the  parameter $\alpha$: ~ for $-\pi/2<\alpha<\pi/2$ the coupling is attractive, whereas for $\pi/2<\alpha<3\pi/2$ it is repulsive. In the first, attractive case the coherent state $\varphi_1=...=\varphi_N$ is always stable (and co-exists with chimera states). It loses its stability only when $\alpha$ crosses $\pm\pi/2$. In the second repulsive case, asymptotically stable is the situation when phase variables $\varphi_i, i=1,...,N,$ are uniformly distributed along the ring. ~ This zero order parameter regime of the Kuramoto model is called \textit{$q$-twisted state}. It is characterized by fixed phase difference $2\pi{q}/N$ between succescive oscillators:~ $\varphi_i=\bar{\omega} t + 2\pi qi/N + C\mbox,\; i=1, \ldots, N$ ~\cite{wsg2006}.  Here  $q$ is a winding  number, an integer that stays for the number of  full twists in  phase around  the ring.   Stability of the $q$-twisted states  for different values of winding number $q$ and the coupling radius $r$ was derived for $\alpha=0$ (the most attractive case) in  ~\cite{wsg2006},  and for $\alpha=\pi$ (the most repulsive case) in  ~\cite{ghm2012}. The stability is preserved for the entirety of the $\alpha$-intervals of attractivity $-\pi/2<\alpha<\pi/2$, and repulsivity  $\pi/2<\alpha<3\pi/2$, respectively [Girnyk, private communication].

Finally, in Ch.4, we study the origin of the chimera states for the repulsive coupling case $\pi/2<\alpha<\pi$. We
show that chimeras grow, as $\alpha$ decreases,  from so-called \textit{multi-twisted states} reported recently in ~\cite{ghm2012}. 
The latter are characterized by phase difference between neighboring oscillators approximately equal $2\pi{q}/N$ in one sector of the ring,
and $-2\pi{q}/N$ in another sector, and they have intermediate values between the two sectors.
Three typical scenarios for the
chimera birth are reported: 1) via inverse saddle-node bifurcation on an invariant curve, also
known as SNIC or SNIPER, 2) via blue-sky catastrophe when two periodic orbits, stable
and saddle, approach each other and annihilate eventually in a saddle-node bifurcation,
and 3) via homoclinic transition, when the unstable manifold of a saddle comes back
crossing the stable manifold giving rise to \textit{Shilnikov homoclinic chaos} ~\cite{sh1965,sh1967,sh1998,sh2001,abs1983,gs1984}.

\section{Cascade of chimera states: attractive coupling}

\subsection{Step-wise coupling}

Without loss of generality we put in Eq.(\ref{model1}) $\omega=0$ and $K=1$ and assume, first,  parameter $\alpha$ belonging to attractive coupling range from $0$ to $\pi/2$. Results of direct numerical simulation of  the model (\ref{model1}) in the two-parameter plane of the phase shift $\alpha$ and the coupling radius $r$  are presented in Fig.\ref{f1}. 

The figure reveals the appearance of regions for chimera states, shown in color, at $\alpha$ close to $\pi/2$ and an intermediate radius of coupling.  
The regions and typical shapes of average frequencies for the chimera states are shown as shaded (colored) tongues and insets, respectively. Alternatively, if phase shift is smaller ($\alpha<1.3$)  or coupling is only local ($r=1/N$) or global ($r=0.5$), chimeras do not exist. In this case the system dynamics is characterized by fully synchronized $q$-twisted behavior states, moreover, the latter is  possible only for $r<0.33$ ~\cite{wsg2006}.

Each chimera state is characterized by the number of intervals of incoherence, so-called \textit{chimera's heads}, where average frequencies  $\bar{\omega}_i$ of individual oscillators $\varphi_i$ (which are actually Poincare winding numbers) appear to be different from the common frequency of synchronized oscillators.  The regions for chimera states with head numbers $h=1,2,4,6,8,$ and $10$ are shown, examples of respective average frequencies are illustrated in the insets. Further decrease of the coupling radius $r$ yields additional high-order regions following an even pair adding cascade $h=12,14,...$. We conclude that only chimeras with even head number $h$ are possible for Kuramoto model with attractive coupling, except of ''classical'' one-headed chimera ($h=1$). 

Detailed structure of high order chimera's regions is illustrated in Fig.\ref{f1}(b). Each region has the shape of a sharp horizontal tongue, width of the regions decreases with increase of head number $h$ (presumably, with some multiplier smaller than 2/3). As one can observe, the regions are packing closer and closer to $r=0$ axis.  Length of the tongues also decreases with increase of $h$ but much slower.  Neighboring chimera's tongues are heavily intersecting, which means that chimeras with different head number $h$ can co-exist. Moreover, the number  of the intersecting sub-regions grows with $h$ too.  E.g., four different type of chimera states coexist in a yellow shaded sub-region in Fig.\ref{f1}(b) denoted  by $h4,h6,h8$, and $h10$.
 
It can be hypothesized that the cascade of the chimera states is accumulated, as $h$ grows, to the corner parameter point $\alpha=\pi/2$, $r=0$.  If so, $\alpha=\pi/2$, $r=0$ is a highly degenerate parameter point where infinitely many multi-headed chimera states co-exist in the same vicinity and their head number $h\rightarrow\infty$. To verify the hypothesis, one have to increase network size $N$.  This,  in turn,  requires significant increase in the calculation time which is proportional to $N^2$.


\begin{figure}[th!]
  \center{\includegraphics[width=1\linewidth]{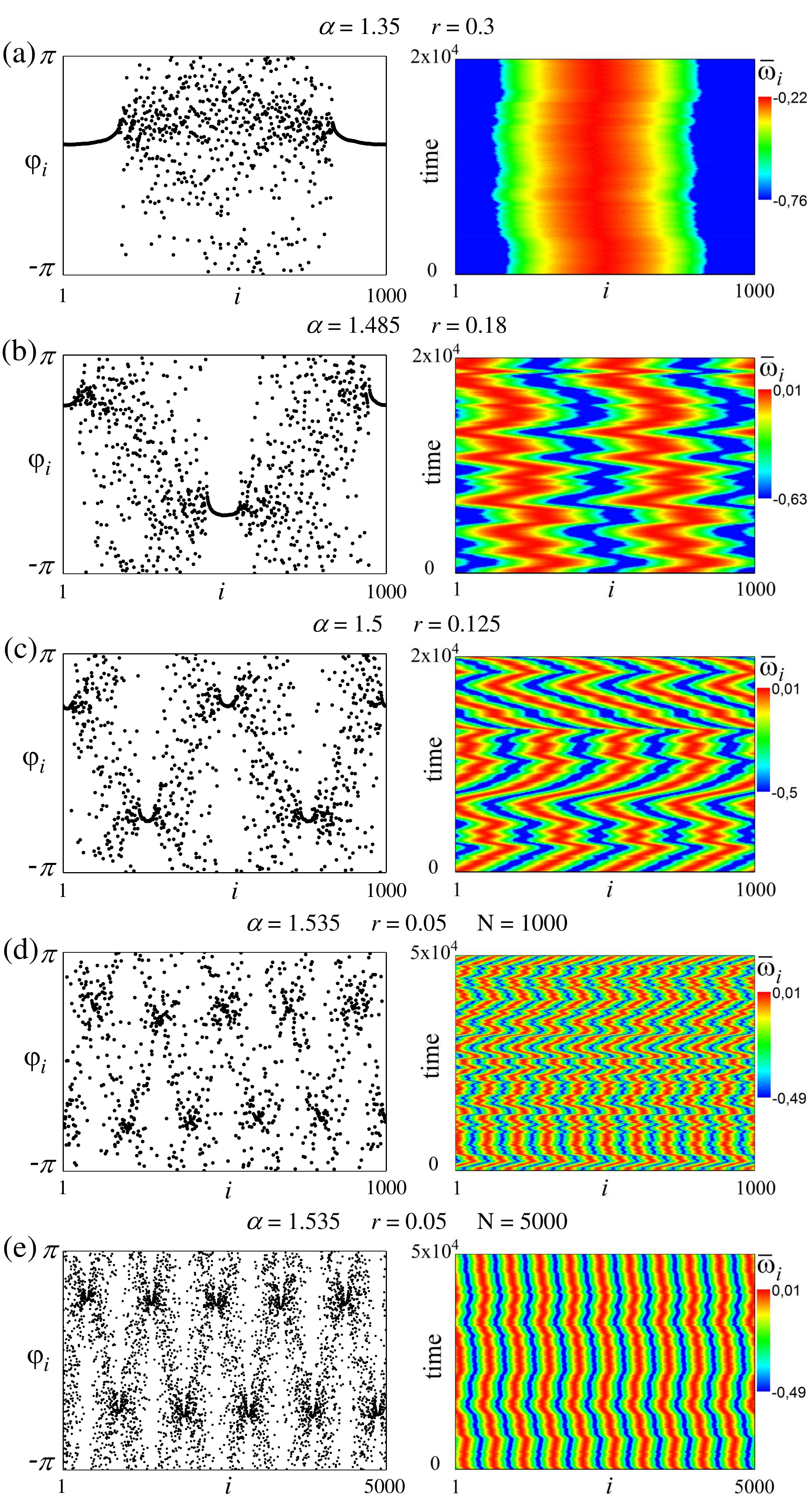}}
 \caption{Examples of multi-headed chimera states  with (a) one, (b) two, (c) four, and (d,e) ten heads (i.e. regions of chaoticity). Snapshots (left columns) and space-time plots of average frequencies (right columns). Other parameters as in Fig.\ref{f1} except of (e) where N=5000.}
  \label{f2}
\end{figure}


Fig.\ref{f2} illustrates typical snapshots of multi-headed chimera states (left column) and respective time courses of their average frequencies (right column), calculated in parameter points $A_1, A_2, A_4$, and $A_{10}$ depicted in Fig.\ref{f1}.  The classical one-headed chimera, Fig.\ref{f2}(a), consists of two oscillator clusters, one coherent and the other incoherent, where the oscillators are synchronized and desynchronized, respectively.  In the coherent cluster all oscillators are nearly in-phase whereas the incoherent clusters oscillators are randomly distributed between $-\pi$ and $\pi$. Moreover, the desynchronized oscillators aren't phase locked, i.e. rotate with different average frequencies (Poincare winding numbers), as it can be concluded from the upper inset in Fig.\ref{f1}(a). The bell-shaped profile of the average frequencies convinces that the oscillators in the incoherent cluster are not synchronized with respect to each other not only in phase but also in frequency. Due to this property, chimera states may be considered as a manifestation of the phenomenon of \textit{partial frequency synchronization}. Indeed, some part of the chimera's oscillators is in perfect synchrony but the others are frequency desynchronized.
The multi-headed chimeras, Fig.\ref{f2}(b-e), are characterized by an even number of the coherent and incoherent clusters alternating successively with each other. The oscillations in the same coherent cluster are nearly in-phase, while between successive 	
coherent clusters they are in anti-phase. 

As it was reported in ~\cite{woym2011}, dynamically, chimeras are high hyperchaotic states in the $N$-dimensional phase state of the model (\ref{model1}),  having a large number of positive Lyapunov exponents. As a consequence, chimera's behavior is significantly irregular in time in spite of its rather regular macroscopic structure in space. This is illustrated by chimera's time courses in Fig.\ref{f2}, right column. Indeed, chimera's position, as one can observe, is not fixed on the ring. On the contrary, each chimera wanders chaotically similar to a Brownian motion ~\cite{owm2010}. Note that chimera's chaotic wandering is a finite-size effect:  wandering amplitude decreases for larger $N$ and eventually, the wandering phenomenon vanishes in the thermodynamic limit $N\rightarrow\infty$ ~\cite{wo2011}.


\begin{figure}[h!]
  \center{\includegraphics[width=1\linewidth]{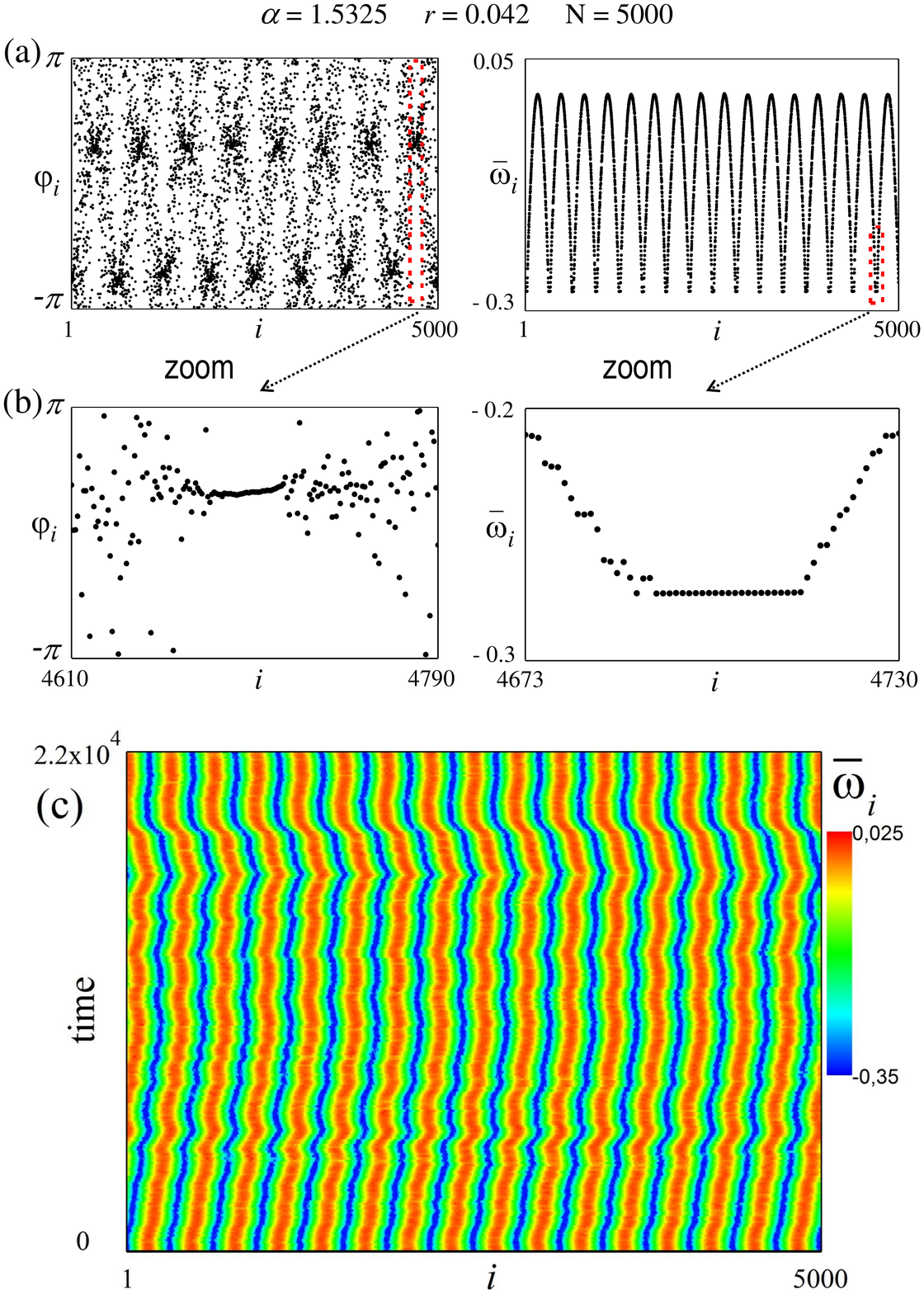}}
 \caption{An example of 16-headed chimera states of the model (\ref{model1}) in parameter point $A_{16}$ shown on Fig.\ref{f1}(b).  Snapshot (left column) and average frequencies of individual oscillators (right column). (b) Enlargement from (a).
  (c) Space-time dynamics. $N=5000$, the other parameters as in Fig.\ref{f1}.}
  \label{f3}
\end{figure}


Surprisingly, as Fig.\ref{f2} also illustrates, wandering amplitude of multi-headed chimeras ($h\geq2$) is much larger than those in one-headed case [compare (b), (c), and (d) with (a)]. The multi-headed chimeras perform long wave chaotic oscillations which are, clearly, different from the standard Brownian motion.  
The nature of the oscillations is not yet understood and should be studied more. 
Note also that chimeras in Fig.\ref{f2} (a-d) are for Kuramoto model (\ref{model1}) of $N=1000$ coupled oscillators. In Fig.\ref{f2}(e) we illustrate changes in the ten-headed chimera's dynamics for $N=5000$ (the other parameters are the same).   As it has been expected, chimera's wandering amplitude is now much smaller.

Fig.\ref{f3} illustrates an example of 16-order chimera state, the highest order we have been able to detect for the model with $N=5000$ oscillators.   Enlargements in  Fig.\ref{f3}(b) confirm perfect  synchronization in a coherent window:  36 oscillators are nearly in-phase [left] and 23 of them are frequency synchronized [right]. Detailed space-time dynamics of the 16-headed chimera state is illustrated in Fig.\ref{f3}(c).

To conclude the discussion for the attractive coupling case of the model (\ref{model1}) with piece-constant coupling function (\ref{Coupling:PW}), we notice one more peculiarity of the chimera's space-temporal dynamics: as one can observe in Figs.~\ref{f2} (left panel) and Fig.\ref{f3}(c), chimeras preserve perfect symmetry in space at any time moment $t$.  Indeed, respective clusters of coherence/incoherence look very much the same, both in length and in color scale.  More precise microscopic inspection argues, nonetheless, that there are minor mismatches but only close to boundaries between the respective coherent/incoherent clusters of the chimera states.

\subsection{Exponential coupling}

Let us now consider the Kuramoto-Sakaguchi model (\ref{model}) with exponential coupling function $G$ given by formula (\ref{modelGexp}). In such a formulation this is exactly the problem,  where chimera state was discovered in the paper ~\cite{kb2002}.  As before, we put $\omega=0$ and $K=1$ and assume parameter $\alpha$ belonging to the attractive coupling range from $0$ to $\pi/2$.

\begin{figure}[h!]
  \center{\includegraphics[width=1\linewidth]{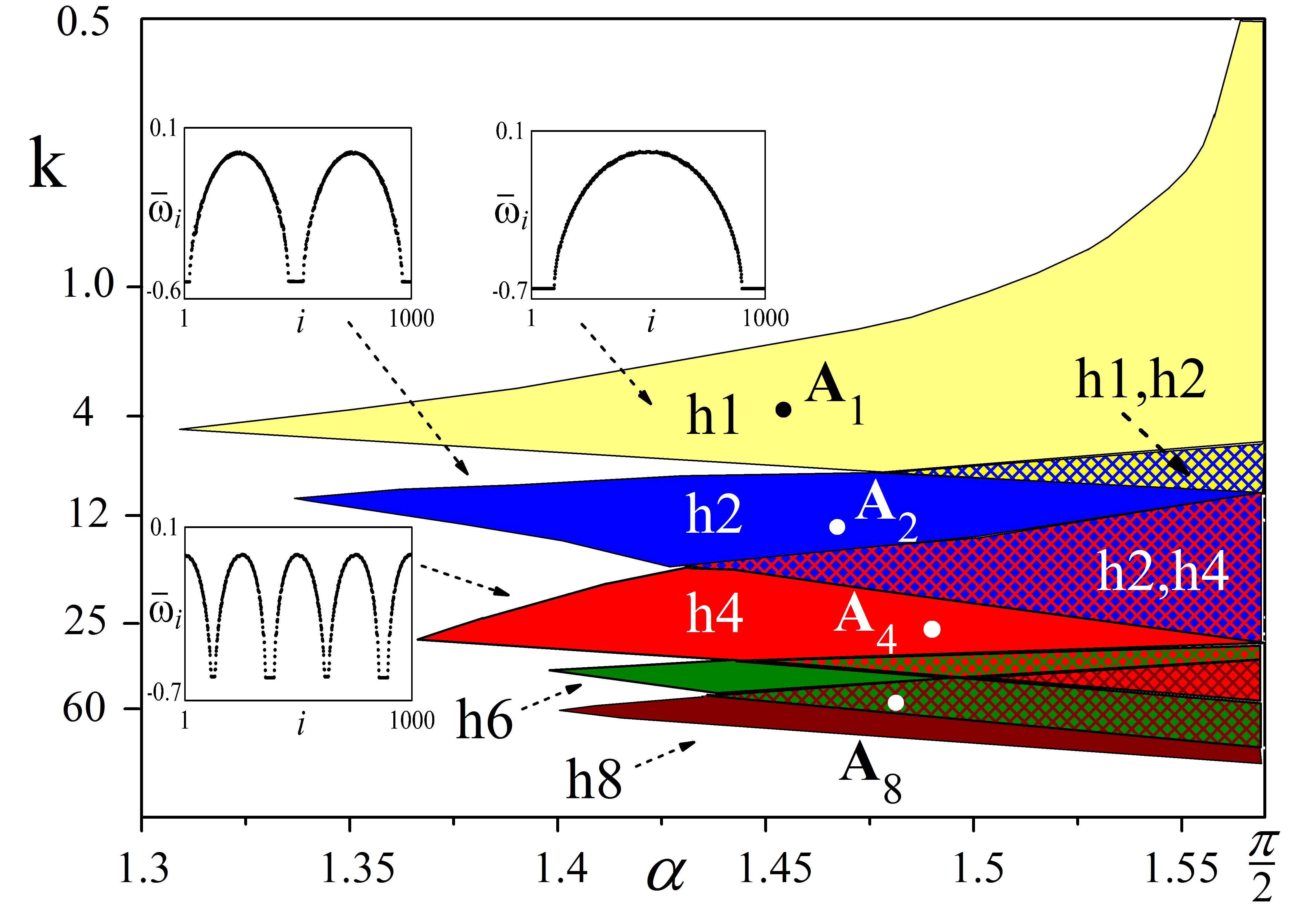}}
 \caption{Regions of multi-headed chimera states in model (\ref{model}) with exponential coupling function (\ref{modelGexp}). Parameter $k$ along vertical axis is displayed using a logarithmic scale. Insets show typical profiles
of average frequencies of individual oscillators in parameter points indicated by arrows. }  
\label{f4}
\end{figure}
\begin{figure}[h!]
  \center{\includegraphics[width=0.98\linewidth]{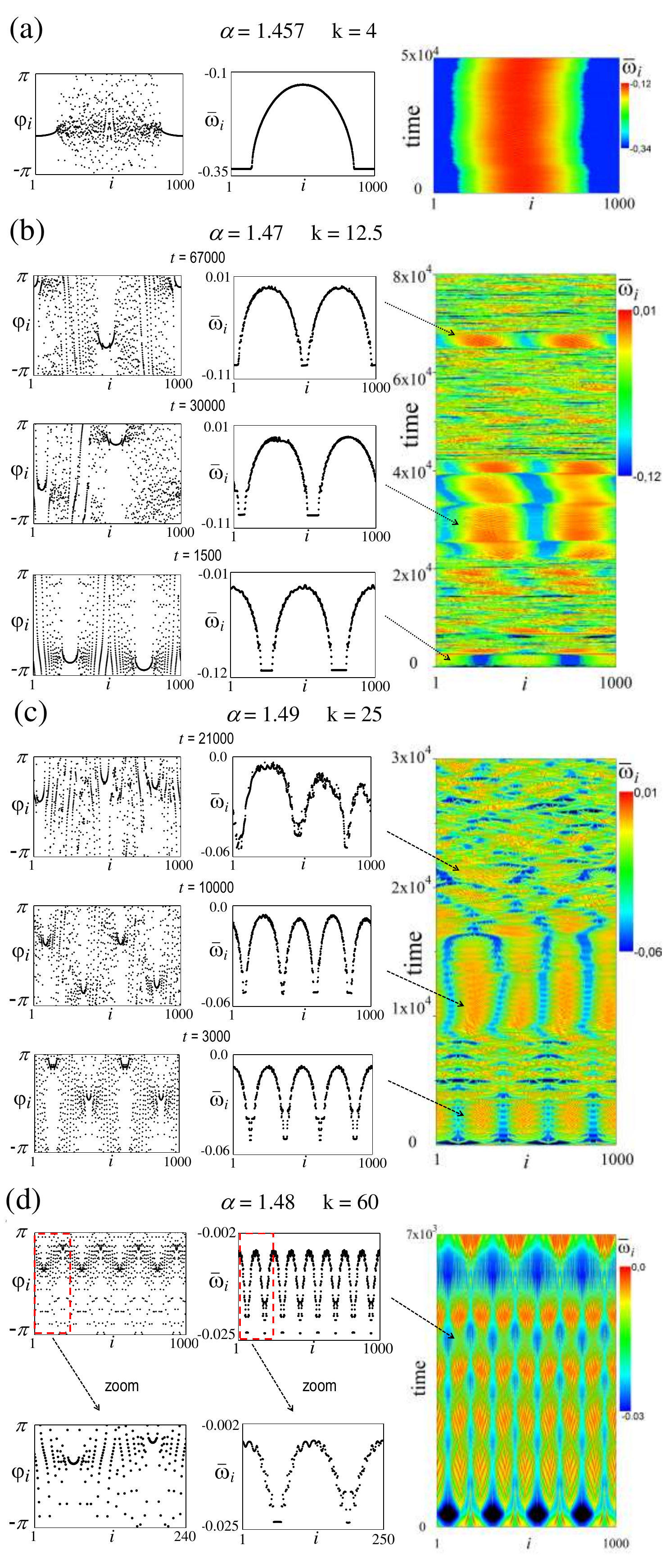}}
 \caption{Examples of multi-headed chimera states in model (\ref{model}) with exponential coupling function (\ref{modelGexp}). Chimera states with (a) one, (b) two, (c) four, and (d) eight heads (intervals of chaoticity) are obtained in respective parameter points $A_1, A_2, A_4$ , and $A_{8}$ depicted in Fig.\ref{f4}. Snapshots (left and middle columns) and space-time plots of average frequencies (right columns). $N=1000,\Delta T=100$.} 
\label{f5}
\end{figure}


Results of direct numerical simulation of  the model (\ref{model}) in the two-parameter plane of the phase shift $\alpha$ and the coupling exponent $k$ (in logarithmic scale) are presented in Fig.\ref{f4}.  The figure is quite similar to figure \ref{f1} for the step-wise coupling case above. It reveals the appearance of regions for chimera states at $\alpha$ close to $\pi/2$ and $k$ between 1 and 80.  As in the step-wise case, the multi-headed chimera regions have the form of sharp horizontal tongues, whose respective average frequencies are also similar, as shown in the inserts.
Fig.\ref{f5} (a) illustrates ''classical'' one-headed chimera state in the parameter point $A_1(\alpha=1.457, k= 4.0)$ reported in ~\cite{kb2002} (cr. Fig.\ref{f2} there).
As one can observe in the right, this state lives for long time, up to $T=50000$ and longer,  performing chaotic wandering of rather small amplitude.
Such a type of behavior is, clearly, similar to what was obtained above in the step-wise coupling case.

On the other hand, to our surprise, all multi-headed chimera states from the regions $h2, h4,..$ in Fig.\ref{f4} are of \textit{transient type} with life time not exceeding several thousand time units. This is illustrated in Fig.\ref{f5} (b-d) where examples of 2-, 4- and 8-headed chimera states are shown.
The right panel shows long term space-time dynamics where chimeras arise only for rather short time interval, then collapse. Sample examples of the transient chimeras, their snapshots and average frequencies, are shown in the left and in the middle panels, respectively. 

All space-time plots in Fig.\ref{f5} are obtained for $N=1000$ oscillators. To overcome the short transient dynamics we tried, following the ideas from ~\cite{wo2011}, to increase the number of oscillators up to 5000 and more. Surprisingly, this has not helped. The multi-headed chimeras preserve the short transient behavior and their life time, virtually, does not change with increase of $N$.  We have no  explanation of the ''short transient phenomenon'', but suspect that it can be connected with global character of the coupling.  Indeed, we can only propose a simple approach how to overcome the short chimera collapsing:  to modify the coupling by introducing a radius $r<0.5$ and not to allow for the exponential coupling to extend above it. Then,  the chimeras live long time as in the  step-wise coupling case described above.


\section{Cascade of chimera states: repulsive coupling}

\subsection{Step-wise coupling}

In this Chapter, we consider Kuramoto model (\ref{model1}) with parameter $\alpha$ belonging to the repulsive range from $\pi/2$ to $\pi$ (assuming as before $K=1$ and $\omega=0$). Results of direct numerical simulation in the two-parameter plane of the phase shift $\alpha$ and the coupling radius $r$  are presented in Fig.\ref{f6}. 
\begin{figure}[h!]
  \center{\includegraphics[width=1\linewidth]{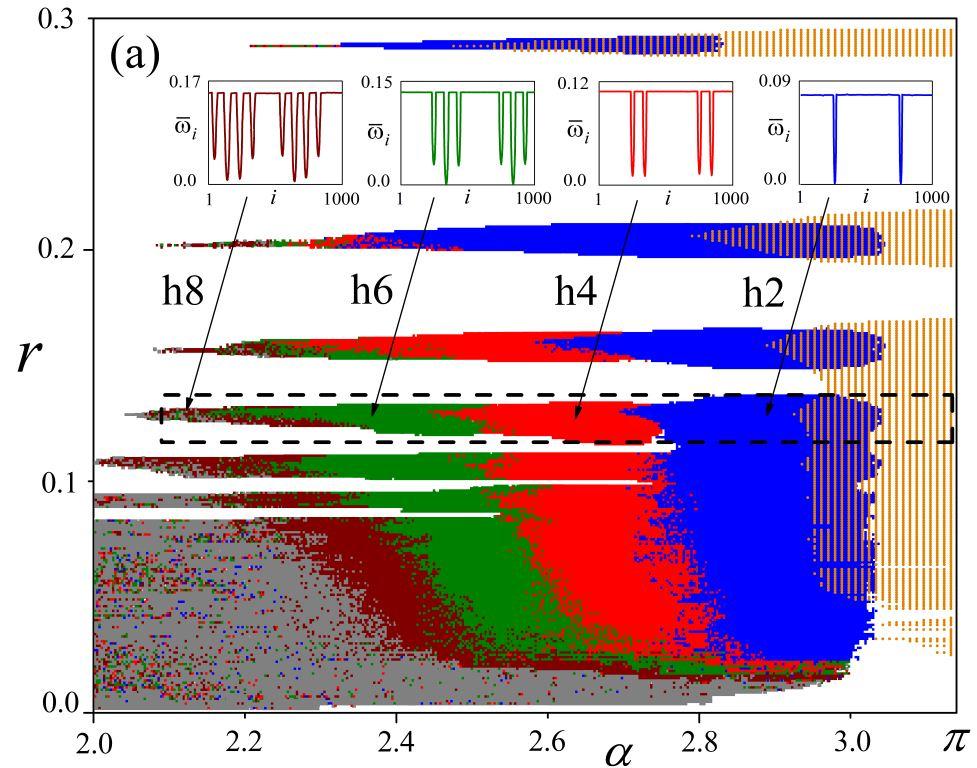}}
  \center{\includegraphics[width=1\linewidth]{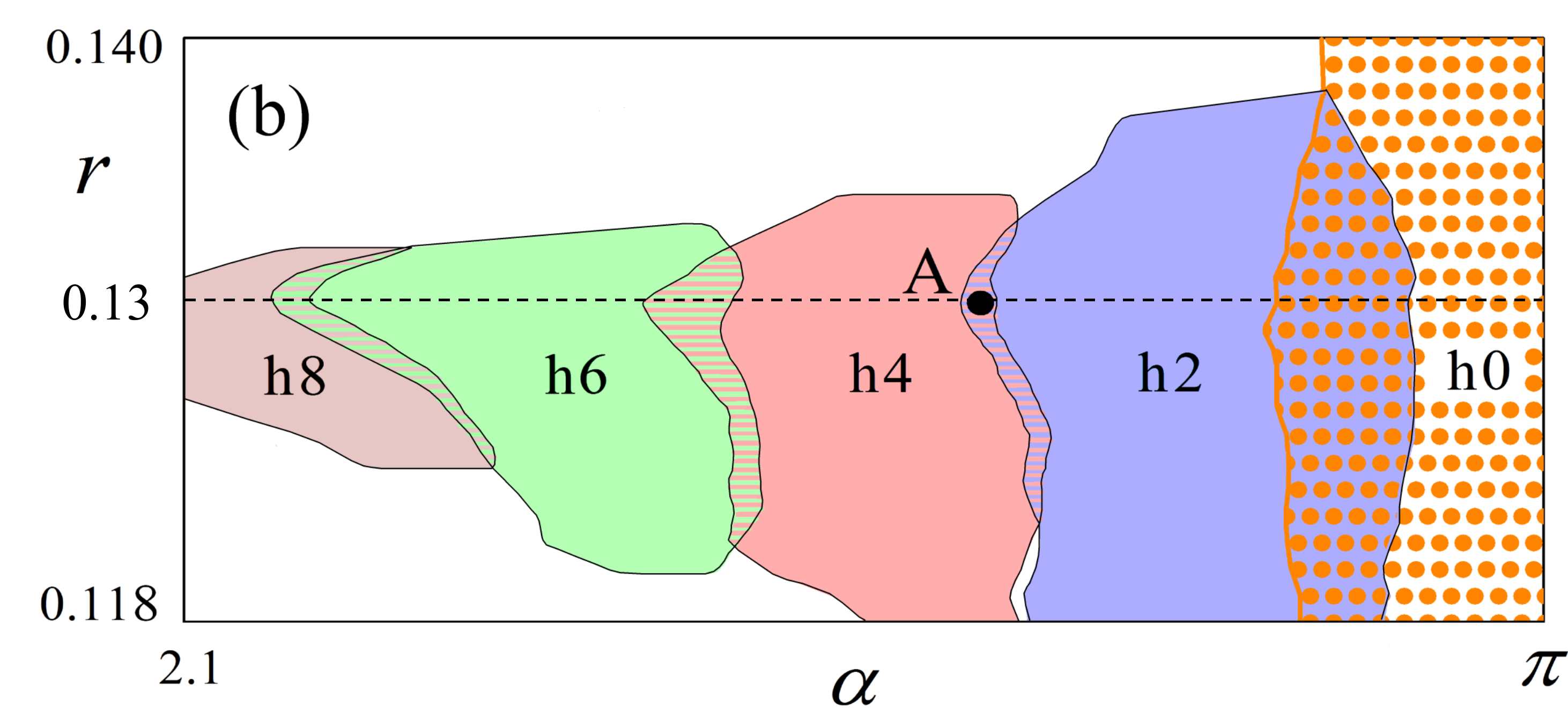}} 
  \center{\includegraphics[width=1\linewidth]{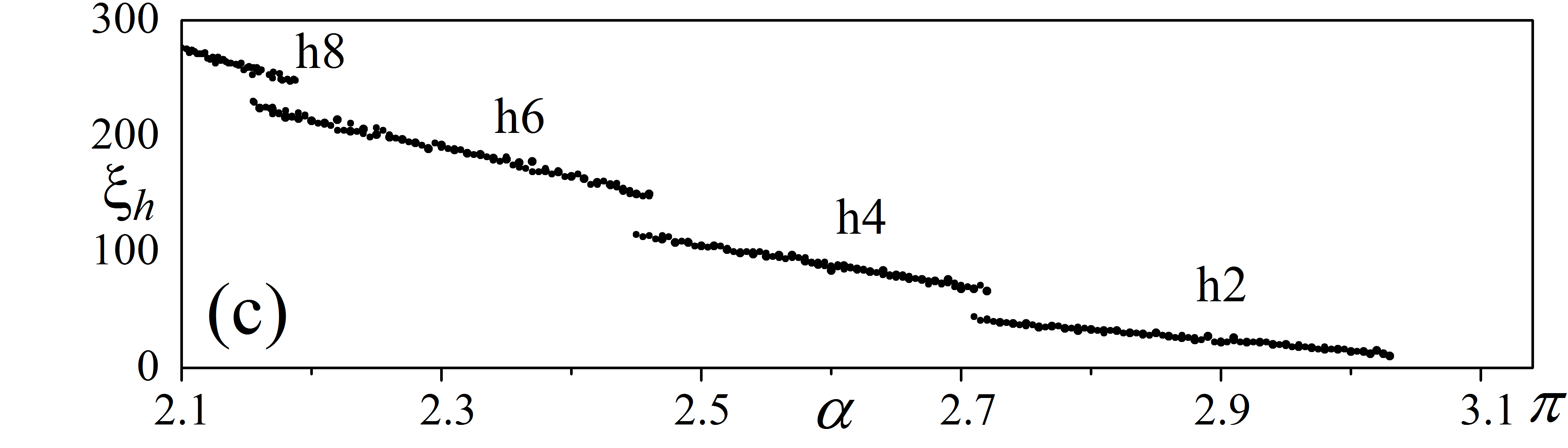}} 
  \caption{(a) Regions of multi-headed  chimera states in model (\ref{model1}) with repulsive coupling, (b) an enlargement of the rectangle from (a), and (c) graph of the number of incoherent oscillators in chimera's heads calculated along the
horizontal line in (b), $N=1000$.}
\label{f6}
\end{figure}


\begin{figure}[h!]
  \center{\includegraphics[width=1\linewidth]{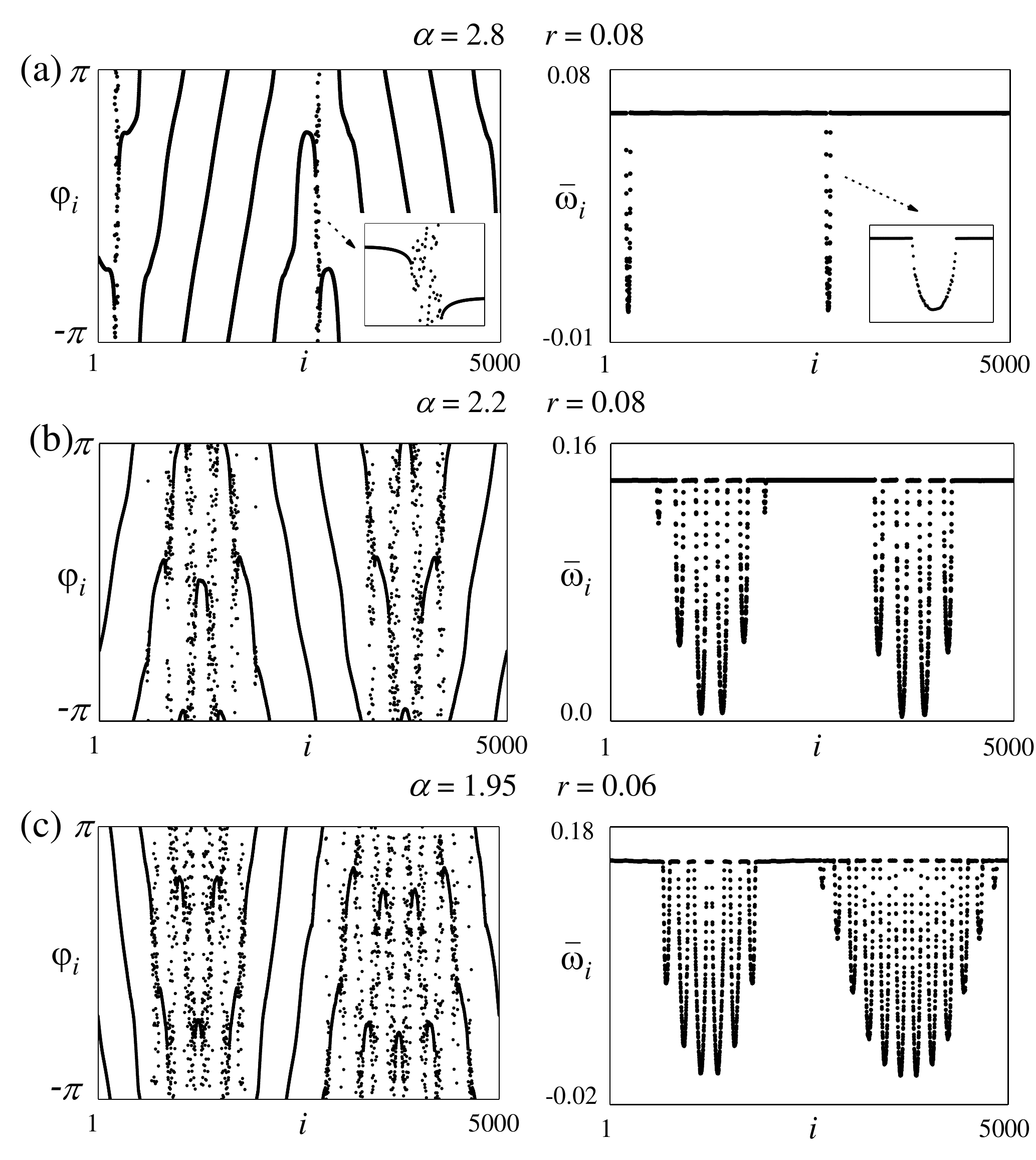}}
  \caption{(a) Examples of multi-headed chimera states for Kuramoto model (\ref{model1}) with repulsive coupling. Snapshots (left panels) and average frequencies (right panels) for the parameter values indicated above the figure. $N=5000$, $\Delta$$T=1000$.}
  
  \label{f7}
\end{figure}


This figure reveals the appearance of regions for chimera states at an intermediate radius of coupling and the phase shift $\alpha$ distanced from  $\pi/2$ and $\pi$.  Alternatively, we observe that if $\alpha$ tends to $\pi/2$ the network displays high-dimensional space-time chaos.  In the opposite situation, when parameter $\alpha$ becomes close to $\pi$ the network  complexity ceases to exist. The  oscillators $\varphi_i$ create so-called \textit{multi-twisted states} ~\cite{ghm2012} which are patterns of non-trivial spatial organization but rotating with a constant velocity, such that they are stationary states in the phase difference variables 
\begin{equation}
\label{newvariables}
\psi_i=\varphi_{i}-\varphi_{i+1},   ~~      i=1, \ldots, N-1.
\end{equation}
In Fig.\ref{f6} (a,b), parameter region of multi-twisted states is dot-hatched in orange, see examples of the states in the Fig.\ref{f10}. Regions of chimera states and typical shapes of the average frequencies are shown in Fig.\ref{f6}(a,b) as shaded (colored) multi-connected vertical strips and respective insets.
A variety of more complicated regimes including traveling waves of different shape and dynamics are detected in our simulations for smaller $r$ and $\alpha$ (respective region is shown in grey), but they are out of the goal of the study. There are also 
$q$-twisted states $\psi_i \equiv 2\pi q/N,  i=1, \ldots, N-1$ co-existing with chimeras and other more complicated states at any $r$ and $\alpha$ in the parameter plane
as shown in Fig.\ref{f6} (a).

We find that each  chimera state in the repulsive parameter region  $\pi/2<\alpha<\pi$ is characterized by even number of narrow layers of incoherence, chimera's heads, where the average frequencies $\bar{\omega}_i$ are different from the frequency of the synchronized clusters.  Only regions with head numbers $h=2,4,6$, and $8$ are shown. Further decrease of $\alpha$ or $r$ yields additional high-order regions following an even pair adding cascade $h=10,12,...$.
The number of incoherent oscillators, i.e., those in chimera's heads, grows with decrease of $\alpha$. This is illustrated in Fig.\ref{f6}(c), where 
we fix the coupling radius $r=0.13$ and vary $\alpha$ along the horizontal line shown in Fig.\ref{f6}(b).

Examples of the multi-headed chimera states, snapshots and shapes of average frequencies, are illustrated in Fig.\ref{f7}. 
Respective space-time dynamics are shown in Fig.\ref{f8}, where one can observe irregular chimera's wandering, as a manifestation of the chaotic system dynamics in the $N$-dimensional toroidal phase space of Eq.(\ref{model},\ref{modelGexp}).


\begin{figure}[h!]
  \center{\includegraphics[width=0.9\linewidth]{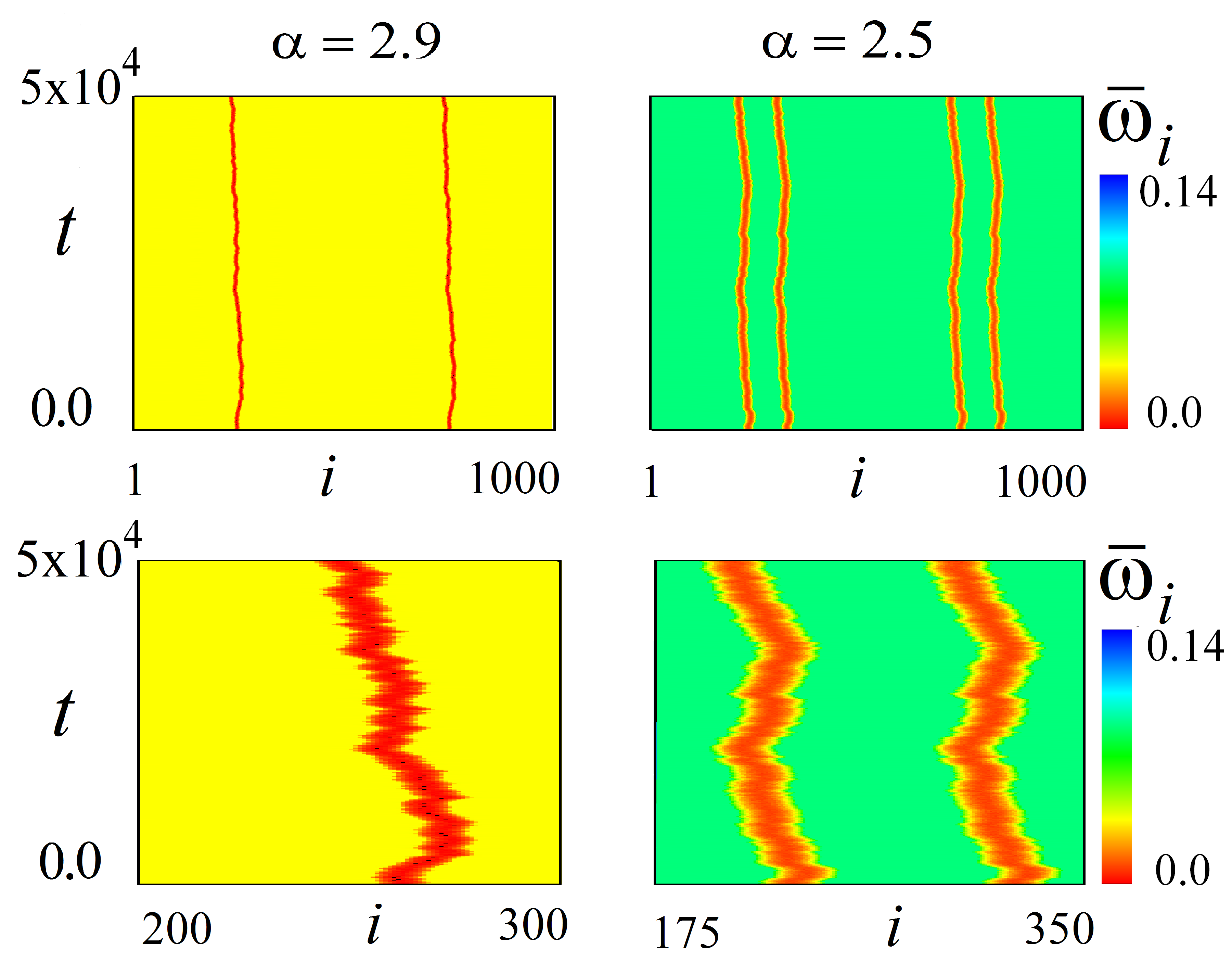}}
  \center{\includegraphics[width=0.9\linewidth]{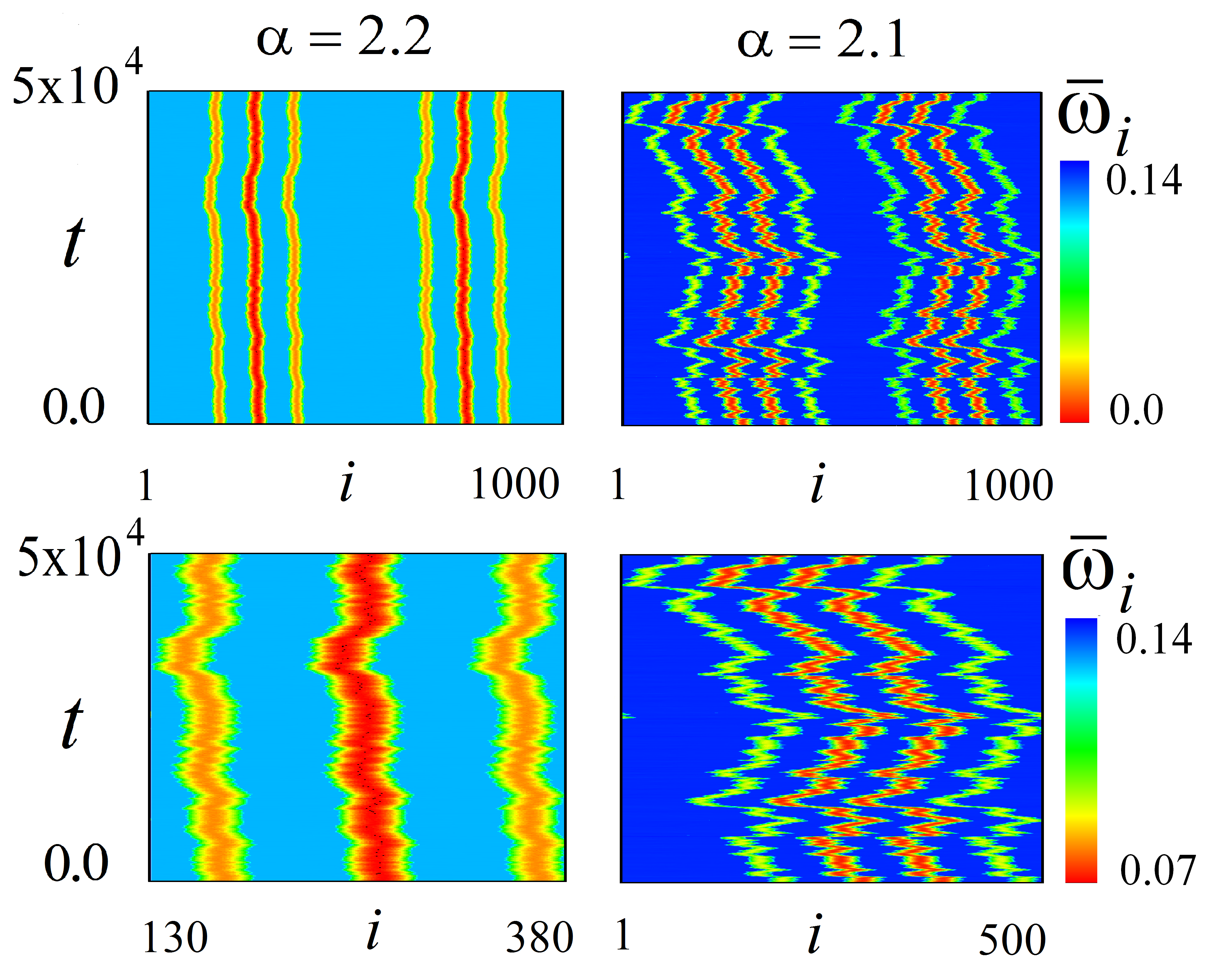}}
  \caption{Examples of space-time dynamics of multi-headed chimera states for model (\ref{model1}) with repulsive coupling. $N=1000, r=0.13,$ $\Delta$$T=100$. }
  
  \label{f8}
\end{figure}


\begin{figure}[h!]
  \center{\includegraphics[width=1\linewidth]{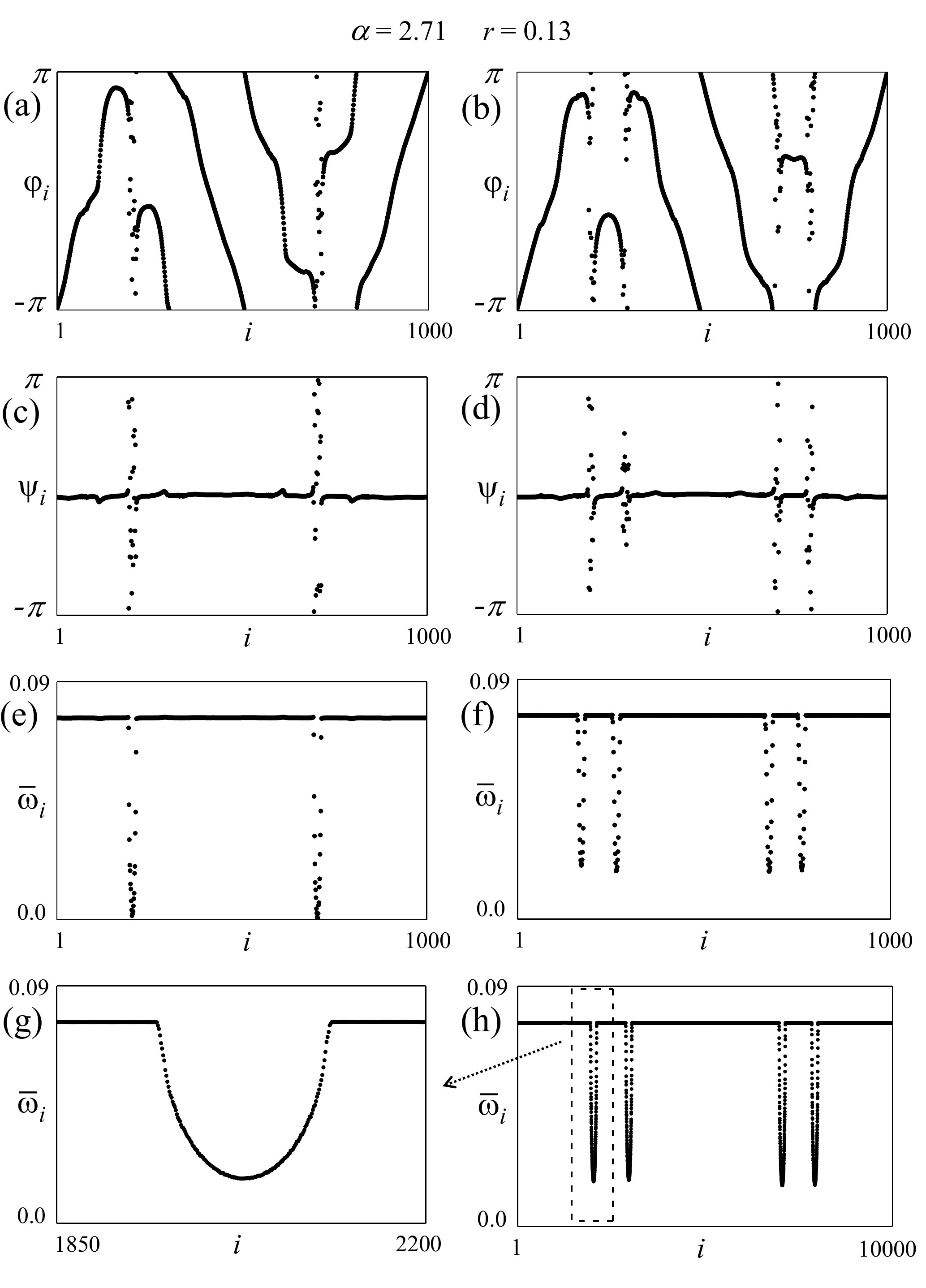}}
  \caption{(a-f) Two co-existing chimera states in parameter point  A for  $\alpha=2.71, r=0.13$ (depicted in Fig.\ref{f6}(b) : two-headed (left) and four-headed (right)  for $N=1000$; (g) four-headed chimera for $N=10000$; (h) enlargement from (g).}
  \label{f9}
\end{figure}

 
\begin{figure}[h!]
  \center{\includegraphics[width=1\linewidth]{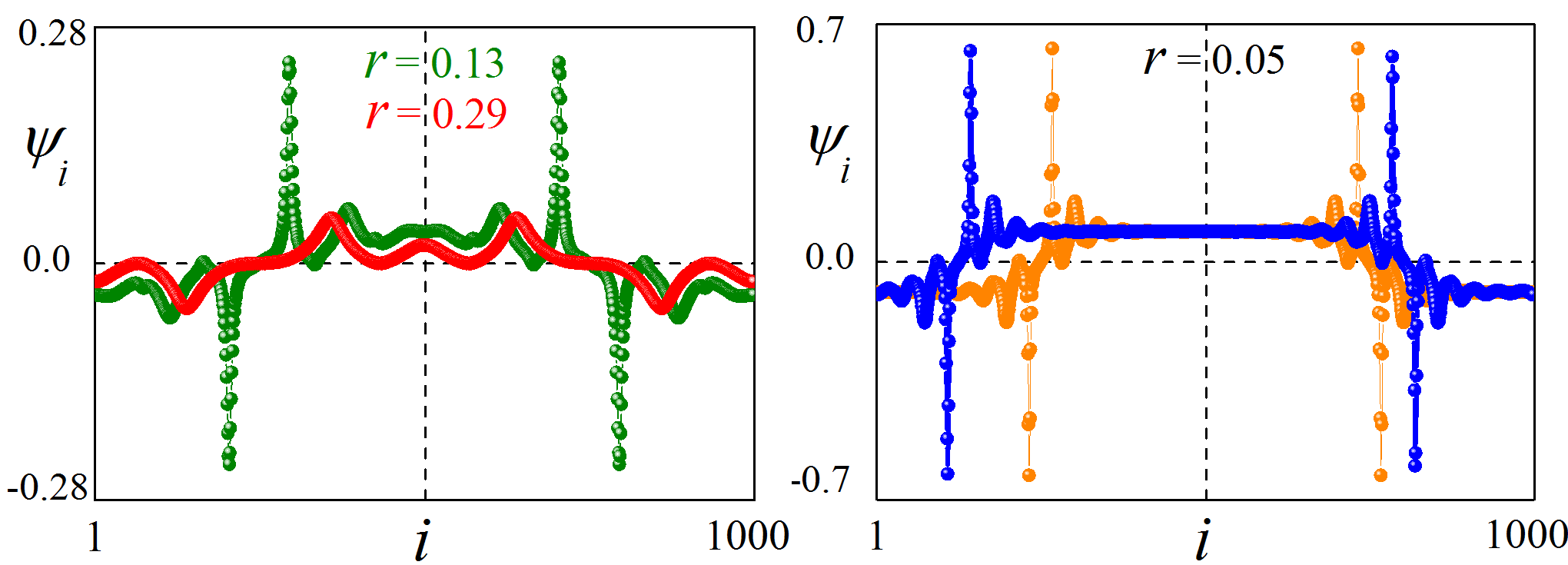}}
  \caption{Examples of multi-twisted states for different $r$. Parameters $N=1000$, $\alpha=\pi$.}
  \label{f10}
\end{figure}

As one can observe in Fig.\ref{f6}(b), neighboring chimera's regions are slightly intersecting and so, chimeras with different head-numbers can co-exist. Fig.\ref{f9} illustrates in details two such co-existing chimeras, with 2 and 4 heads respectively, in parameter point $A$ depicted in Fig.\ref{f6}(b).  

We find that chimeras are born, as $\alpha$ decreases, developing from the multi-twisted states of the form illustrated in Fig.\ref{f10}.   Only blue state is not symmetric, its continuation  (with decrease of $\alpha$) produces a chimera state at the boundary of the blue region in Fig.\ref{f6}(a). Three others, symmetric multi-twisted states survive up to the left boundary of orange region, where they disappear in a crisis without any chimera birth. Just after the birth, only one or few oscillators split up from the main synchronized cluster and start to rotate with  different average frequencies (Poincare winding numbers).  With further decrease of $\alpha$ additional oscillators split up so that incoherence strips, chimera's heads, become wider, see Fig.\ref{f6} (c).  Eventually, chimera states cease to exist at the left hand boundary of the regions in a crisis bifurcation. 

Dynamical complexity of the chimera state can be characterized by the number of positive Lyapunov exponents or, more precisely, by Lyapunov dimension.  Our simulations confirm, in analogy with the case of attractively coupled Kuramoto model ~\cite{woym2011}, that Lyapunov dimension is slightly smaller than the number of incoherent oscillators.  Therefore, the complexity grows with decreasing $\alpha$ as it is confirmed by Fig.\ref{f6}(c) and by the graph of 80 largest Lyapunov exponents  in Fig.\ref{f11} (N=1000). For $\alpha<2.45..$ more than 80 Lyapunov exponents become positive.  At $\alpha=2.71$ both chimeras, two-headed and four-headed, shown in Fig.\ref{f9} contain $\approx{20}$  incoherent oscillators in each head. Then, our simulations show that 2-headed chimera has 17  positive Lyapunov exponents with Lyapunov dimension $D_L\approx{36.3}$, 4-headed chimera has 28 positive Lyapunov exponents with Lyapunov dimension $D_L\approx{59.4}.$ 

\begin{figure}[ht!]
  \center{\includegraphics[width=1\linewidth]{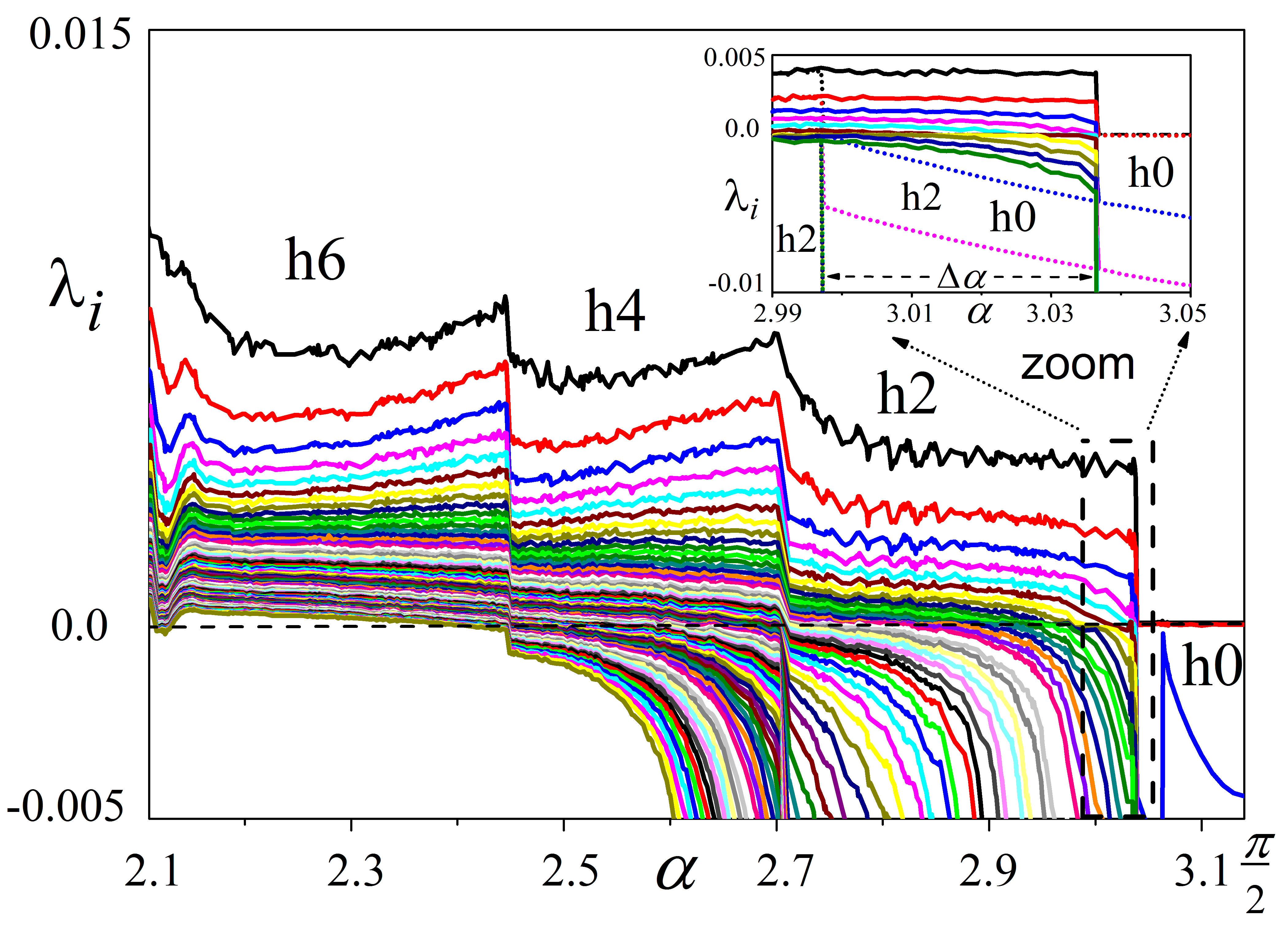}}
  \caption{ Graph of 80 largest Lyapunov exponents  
  computed for chimera trajectories of model (\ref{model1}) with coupling radius $r=0.13$, $N=1000$ and step $h_{\alpha}=0.001$. For each value of $\alpha$, trajectory length  $T=3$x$10^{5}$. The calculations were performed at the computer cluster ''CHIMERA'' (http://nll.biomed.kiev.ua/cluster) and GRID infrastructure ~\cite{zssb2007}. Enlargement in inset: Ten largest LEs of two-headed chimera states (solid lines) and multi-twisted state (dotted lines) co-existing in the parameter interval $\Delta \alpha$.} 
    \label{f11}
\end{figure}

 In the thermodynamic limit $N\rightarrow\infty$, the number of incoherent oscillators grows proportionally to $N$. The limiting profile of average frequencies is illustrated in Fig.\ref{f9}(g,h), it can be obtained by standard self-consistency approach developed in ~\cite{kb2002,as2004,bpr2010,woym2011}.  The profile is obtained numerically  for $N=10000$ oscillators, it is similar but 10 times disperser compared to the respective $N=1000$ profile shown above in Fig.\ref{f9}(e,f).

\section{Appearance of the chimera states: three scenarios}

We find three typical scenarios for the chimera birth: 1) via inverse saddle-node bifurcation on a closed invariant curve knowN also as SNIC or SNIPER [Fig.\ref{f12}]; 2) via blue-sky catastrophe when two periodic orbits, a limit cycle and a saddle cycle, approach each other collapsing in a saddle-node periodic orbit [Fig.\ref{f13}]; and 3) via homoclinic transition when the phase slips (producing actually a chimera) arise as a destabilization of a two-loops limit cycle born in a neighborhood of a homoclinic trajectory [Fig.\ref{f14}].
To unfold the transitions, we fix the coupling radius $r$ and decrease the phase shift $\alpha$ from $\pi$.  In each of the cases, as we have found, chimera state grows from multi-twisted states of the form illustrated in Fig.\ref{f10}. For the convenience, below we use new phase difference variables 
\begin{equation}
\label{verynewvariables}
\theta_i=\varphi_{i+1}-\varphi_1,~~ i=1, \ldots, N-1.
\end{equation}

In Fig.\ref{f12},  we take model (\ref{model1}) with $N=54$ oscillators, each coupled to $P=8$ nearest neighbors, then the coupling radius $r=P/N$ equals $0.144...$ . Then chimera state grows, with decreasing $\alpha$, from a pair of multi-twisted states which are similar but shifted with respect to each other by one oscillator, see Fig.\ref{f12}(b).  


\begin{figure}[h!]
  \center{\includegraphics[width=1\linewidth]{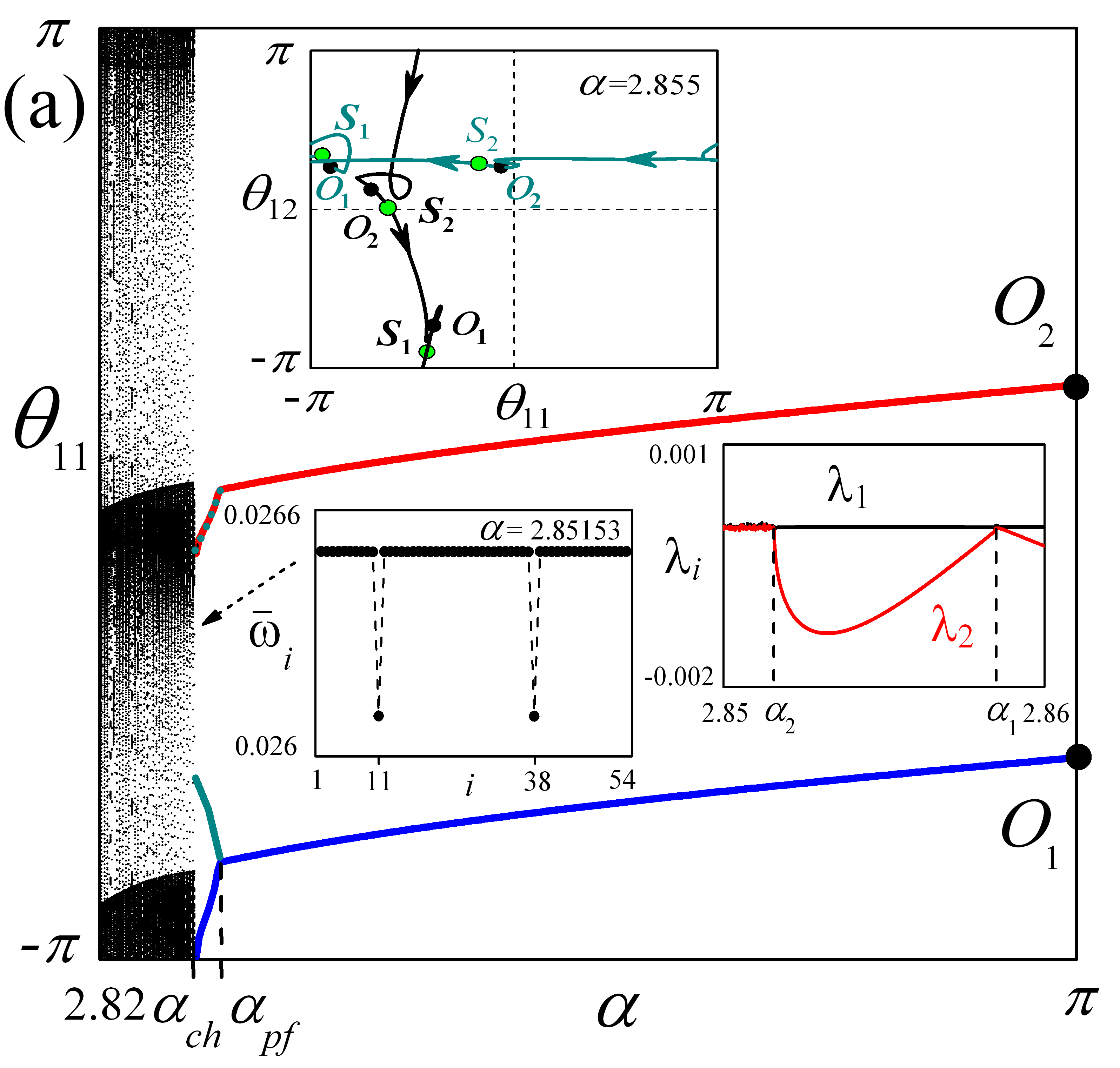}}
  \center{\includegraphics[width=1\linewidth]{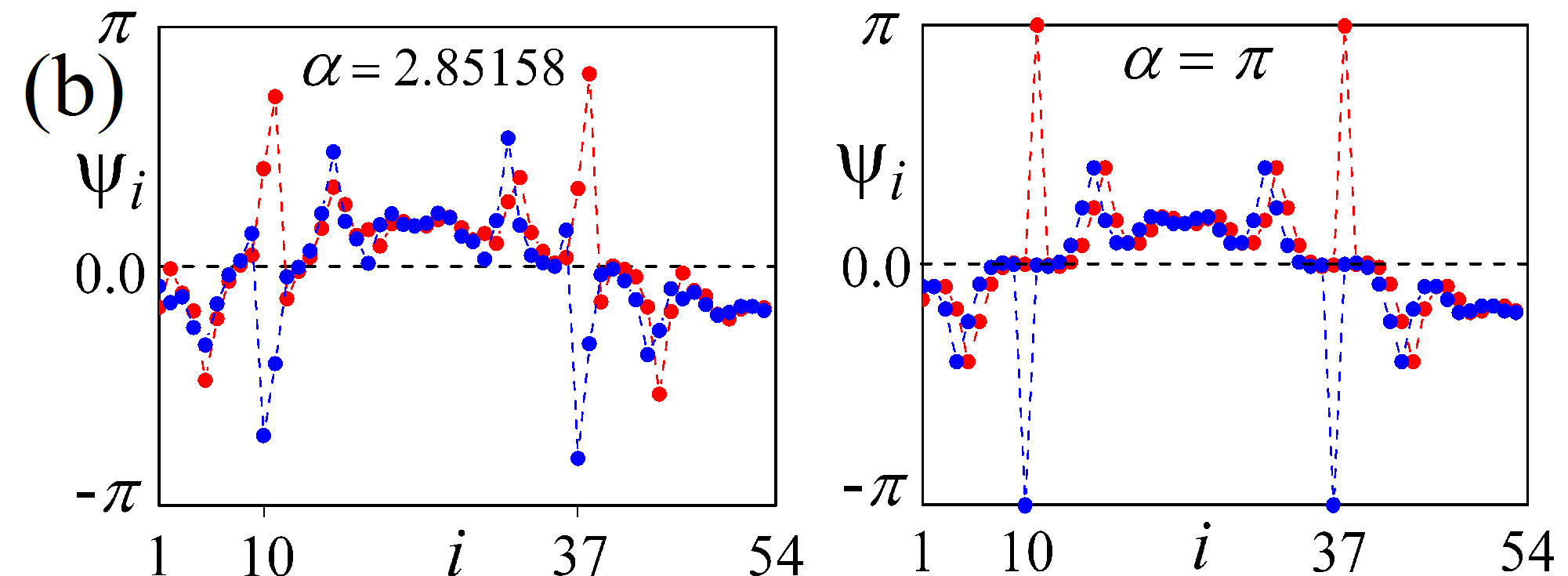}}
  \caption{Saddle-node scenario, see text for details.}
  \label{f12}
\end{figure}

Both multi-twisted states obey an axis symmetry and have two oscillators staying in anti-phase to the others (shown in Fig.\ref{f12}(b) for $\alpha=\pi$). They are $\varphi_{10}$ and $\varphi_{37}$ for one of the multi-twisted states (shown in blue) and $\varphi_{11}$ and $\varphi_{38}$ for the other (shown in red).  The anti-phase oscillators are playing role of a ''germ'' for the chimera state. Indeed, they just split up and start rotating in a chimera birth bifurcation at $\alpha_{ch}=2.85855...$.  In Fig.\ref{f12}(a) the multi-twisted states are depicted as $O_1$ and $O_2$. On the way from $\alpha=\pi$  to $\alpha_{ch}$ they undergo a pitchfork bifurcation at some $\alpha_{pf}$, each producing a pair of asymmetric states of a shape shown in Fig.\ref{f12}(b).  With further decrease of $\alpha$, two of the states (by one from each pair) find themselves  at an invariant resonant curve, as it is shown in a projection in upper inset in Fig.\ref{f12}(a). The chimera-birth bifurcation takes place in the next moment $\alpha_{ch}$, when the states simultaneously disappear in a collision with respective saddles. The bifurcation produces slips of oscillators $\varphi_{11}$ and $\varphi_{38}$ around the circle, first very slow then faster and faster with more decrease of $\alpha$.  The same type of the behavior takes place for the symmetrically located oscillator $\varphi_{38}$.  A two-headed chimera state is born, its average frequency characteristic is shown in the lower inset in Fig.\ref{f12}(a).

\begin{figure}[h!]
  \center{\includegraphics[width=1\linewidth]{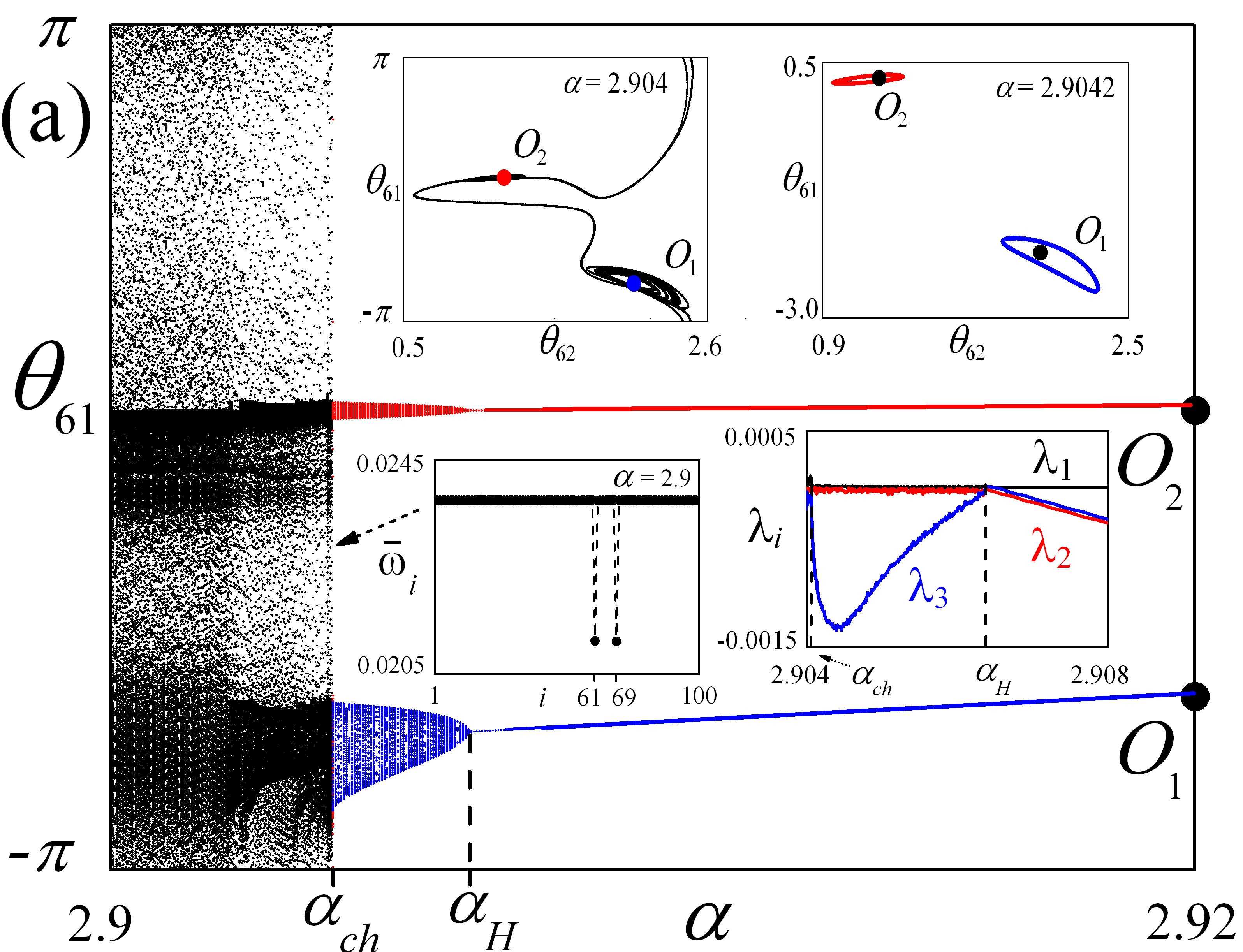}}
  \center{\includegraphics[width=1\linewidth]{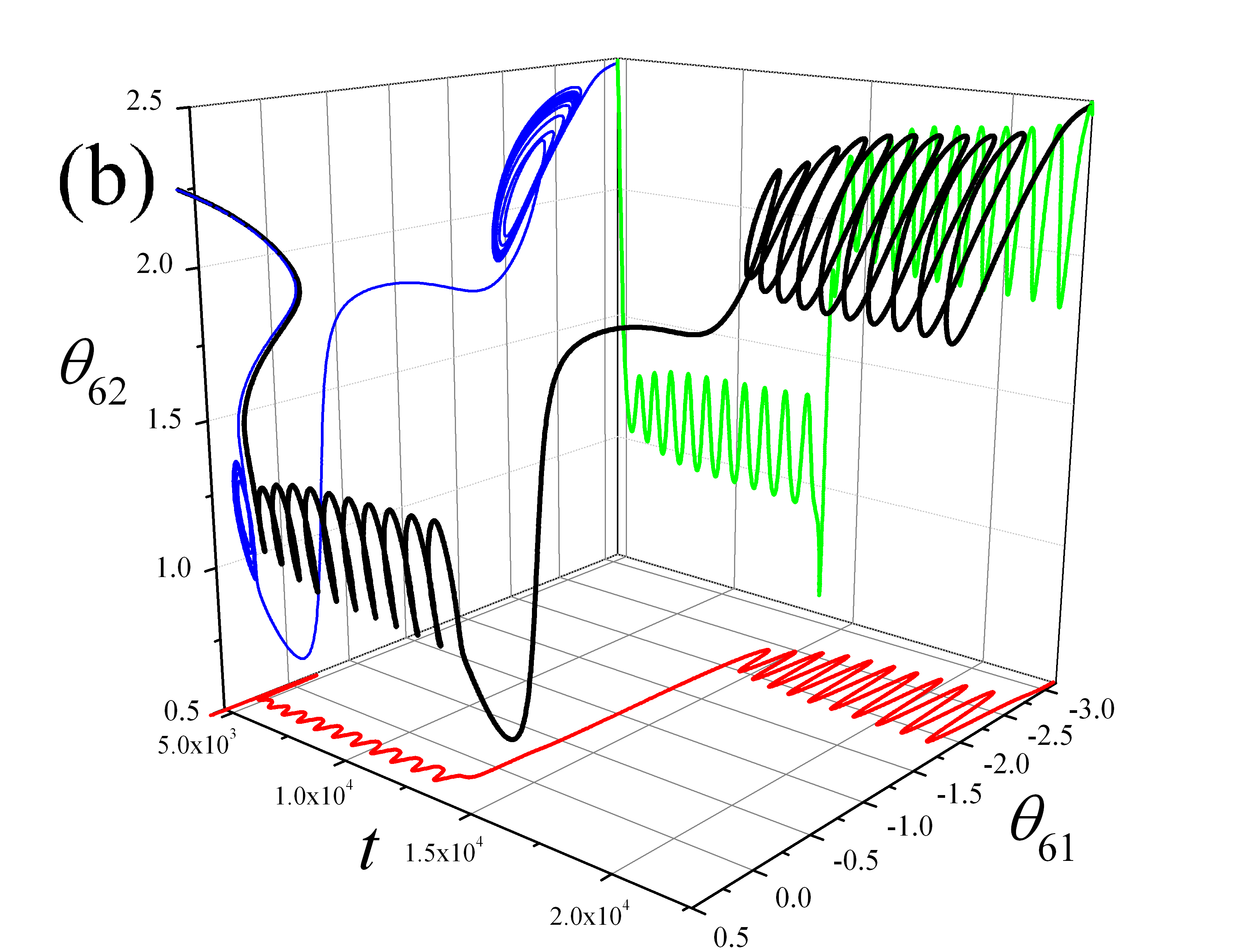}}
  \caption{Blue-sky catastrophe scenario, see text for details.}
  \label{f13}
\end{figure}
The second, \textit{blue-sky catastrophe} scenario is illustrated in Fig.\ref{f13}. Here $N=100$ and $P=10$ (coupling radius $r=0.1$). A difference from the previous scenario is that circular rotations along an invariant curve arise not due to a saddle-node collapse but as a collision and annihilation of two cycles (periodic orbits), one stable and the other saddle. This is illustrated in a projection in upper left insets in Fig.\ref{f13}(a).  In other words, the chimera birth bifurcation represents again a kind of an inverse saddle-node at a resonant invariant curve at $N$Dim torus, but for periodic orbits, not for equilibrium, which is known as blue-sky catastrophe ~\cite{pp1974,sh1998,sh2001}. We find that for the chosen parameters, the bifurcation occurs at $\alpha_{ch}= 2.9041...$: chimera state is born from two limit cycles, shown in right upper insets.  The cycles, in turn, have been born in a Hopf bifurcation at $\alpha_H=2.9067...$ from the multi-twisted states  Fig.\ref{f13}(a).   As in the previous saddle-node scenario, two multi-twisted states participate in the chimera creation, they are depicted by $O_1$ and $O_2$ in Fig.\ref{f13}(a).  The multi-twisted states are similar but shifted  along the ring by 7 oscillators. The shifting is reflected, eventually, in the appearing chimera state structure, where just two oscillators $\varphi_{61}$ and $\varphi_{69}$ rotate differently from the others, see lower inset. Furthermore, Fig.\ref{f13}(b) illuminates 3Dim chimera state dynamics just after the blue-sky catastrophe bifurcation at $\alpha =2.90407..$.
 

\begin{figure}[h!]
  \center{\includegraphics[width=1\linewidth]{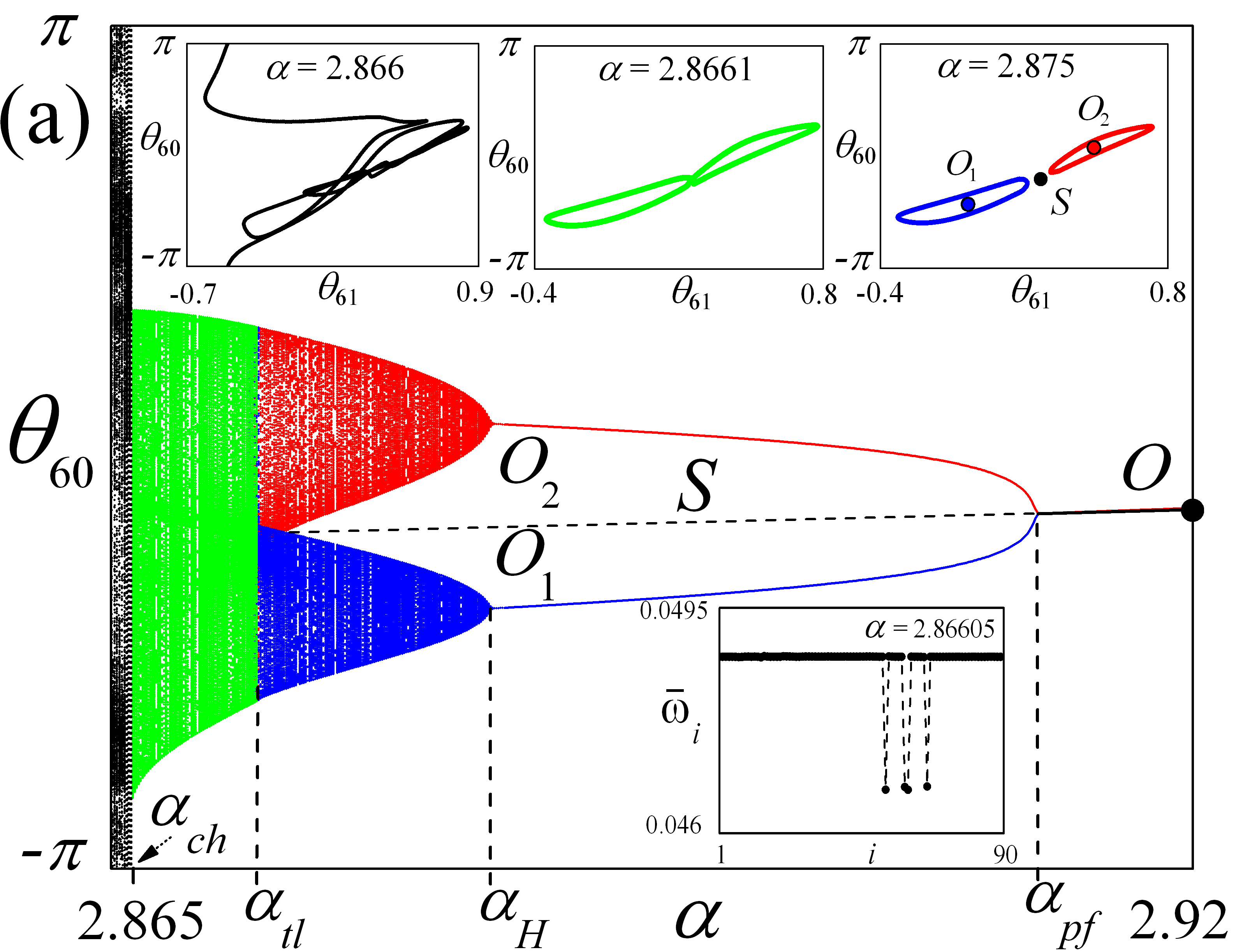}}
  \center{\includegraphics[width=1\linewidth]{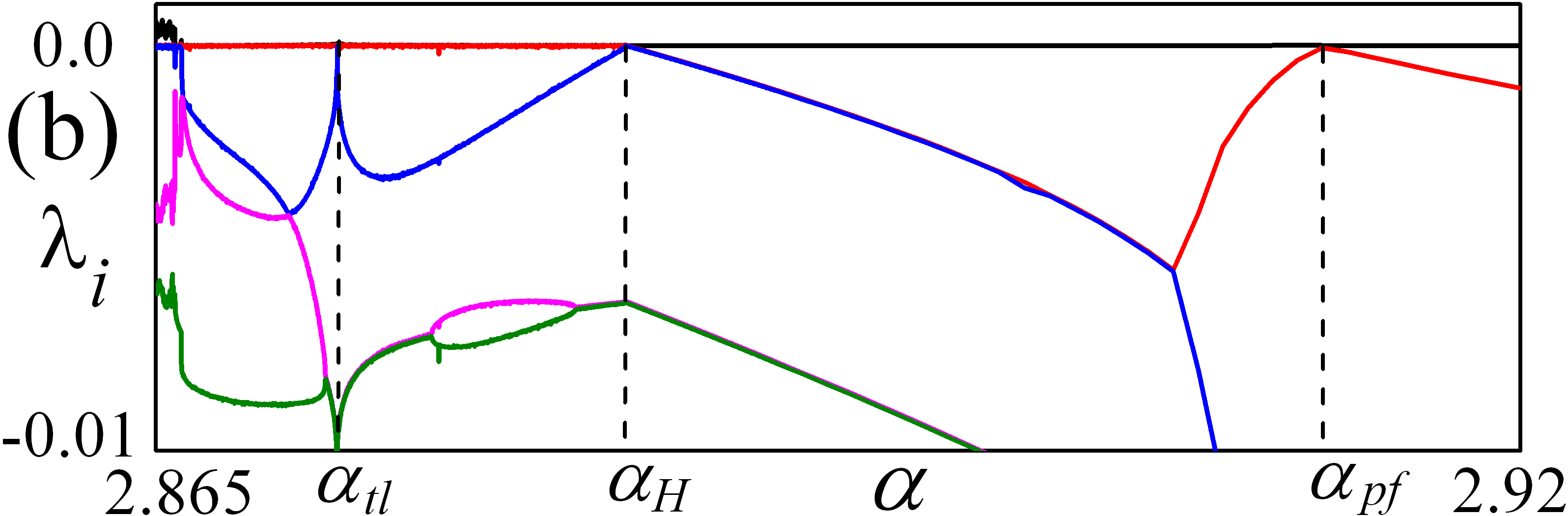}}
  \center{\includegraphics[width=1\linewidth]{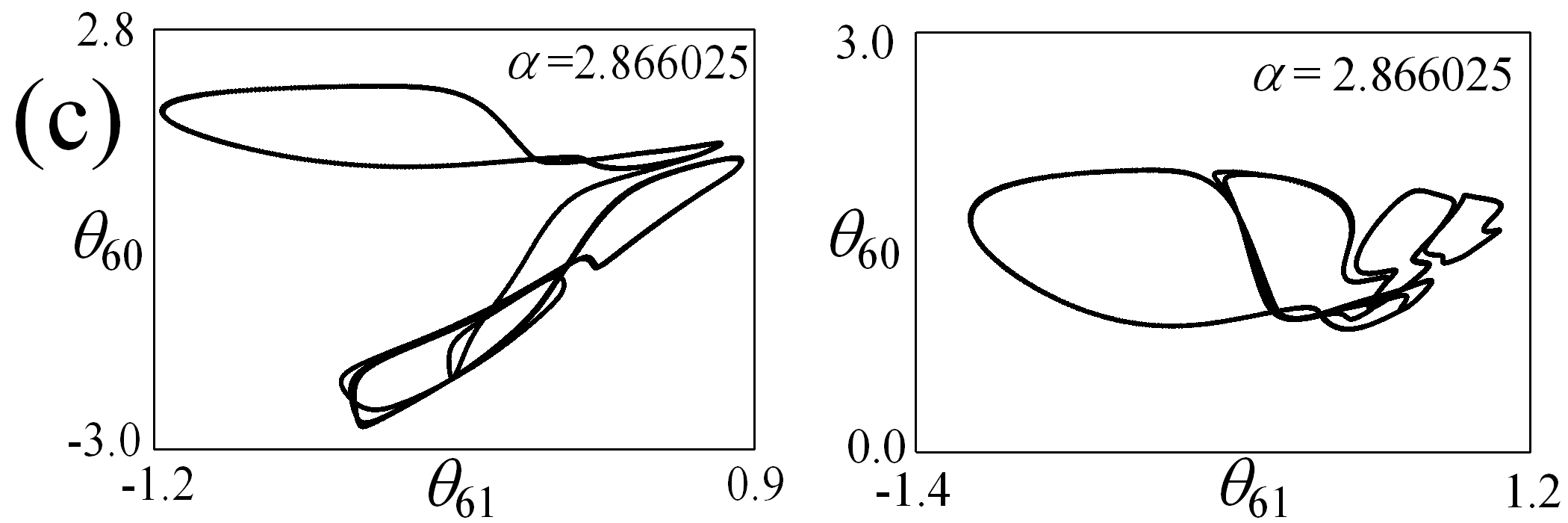}}
  \caption{Homoclinic scenario, see text for details.}
  \label{f14}
\end{figure}
 
The third, \textit{homoclinic scenario} [Fig.\ref{f14}, $N=90, r=0.1$] differs from the others from the very beginning. Chimera state in this case is originated from a unique, symmetric  multi-twisted state. At $\alpha_{pf}=2.91215...$ it undergoes a pitchfork bifurcation producing two asymmetric states which, in turn, give birth to two stable periodic orbits in a Hopf bifurcation at $\alpha_H=2.88441...$. With further decrease of $\alpha$ both periodic orbits grow in size and move to each other [Fig.\ref{f14}(a),~upper right inset]. The cycles are symmetrically located with respect of a saddle state $S$ (which is a continuation of the original multitwisted state $O$), and they are close to creating a double ''eight-like'' homoclinic contour.  But, as we have concluded after more detailed simulations, this does not happen. Instead, as we assume, homoclinic bifurcation of the saddle cycle $S$ takes place in a different way at some $\alpha$ before. It results in complex dynamics, including Smale horseshoe, in phase space around $S$. Therefore, both cycles disappear not in a symmetric homoclinic bifurcation but in an inverse saddle-node bifurcation, at $\alpha_{cr}=2.86608...$ [Fig.\ref{f14}(a)]. Slightly before, at $\alpha_{tl}=2.87244...$, a two-loops cycle is born in another saddle-node bifurcation. Cycles of both types, one-loop and two-loops, co-exist in the $\alpha$-range between these two values $\alpha_{tl}$ and $\alpha_{cr}$.

Chimera state arises from a two-loop cycle at some $\alpha_{ch}$, when the cycle loses stability and starts producing slips [$\theta_{60}$ in upper left inset ]. Simultaneously, three more oscillators, $\theta_{53}, \theta_{59},\theta_{66}$, start rotating as it can be concluded from the chimera's frequency characteristic [lower right inset].  The chimera is born. In Fig.\ref{f14}(c), one can also see a 4-loops cycle [left] and a more complicated periodic orbit [right], both coexisting with the chimera state [Fig.\ref{f14}(a), upper left inset] and the 2-loops cycle [Fig.\ref{f14}(a), upper middle inset] at the same parameter value $\alpha=2.866025$.  The system dynamics becomes more and more complicated and multistable. We believe that this complexity is a sign of \textit{Shilnikov homoclinic chaos}  ~\cite{sh1965,sh1967,sh1998,sh2001,abs1983,gs1984}.

These three scenarios do not cover all transitions to the chimera states, they just illustrate how rich and amazing  the dynamics of coupled oscillators could be even if the oscillators are all identical and coupling scheme in the network is symmetric. Indeed, the complexity arises from the internal nonlinear interactions, not from the external signals.  One challenging problem is the question of which of the unique properties of the model are robust and are preserved under perturbations, whether random or regular.

\nonumsection{Acknowledgments} \noindent We thank Matthias Wolfrum and Martin Hasler for illuminating, helpful discussions. We also thank
the Ukrainian Grid Infrastructure for providing the computing
cluster resources and the parallel and distributed software used
during this work. This work was supported
by bilateral German-Ukrainian DFG grant WO 891/3-1.

\end{twocolumn}
\end{document}